\documentclass[11pt, longbibliography]{revtex4-2}

\usepackage[hmargin=1in, vmargin=1in]{geometry} 
\usepackage[utf8]{inputenc}
\usepackage[T1]{fontenc}
\usepackage{upgreek, commath}
\usepackage{listings}
\usepackage{siunitx}

\newcommand{\bl}{\begin{flalign}}
	\newcommand{\enl}{\end{flalign}}

\newcommand{\mc}[1]{\mathcal{#1}}

\newcommand{\tdse}{time-dependent Schr\"{o}dinger equation}

\newcommand{\half}{\frac{1}{2}}
\newcommand{\intf}{\int_{-\infty}^{\infty}}
\newcommand{\inth}{\int_0^\infty}

\renewcommand{\bf}[1]{\mathbf{#1}}

\newcommand{\pos}{{(+)}}
\renewcommand{\neg}{{(-)}}

\usepackage{graphicx}
\usepackage{amssymb}
\usepackage{epstopdf}
\usepackage{amsmath,dsfont}
\usepackage{bm}
\usepackage{braket}
\usepackage[colorlinks,citecolor=blue,linkcolor=blue]{hyperref}
\usepackage{tikz-feynhand}
\usepackage{changes}

\newcommand{\be}{\begin{equation}}
	\newcommand{\ee}{\end{equation}}
\newcommand{\bea}{\begin{eqnarray}}
	\newcommand{\eea}{\end{eqnarray}}
\newcommand{\ba}{\begin{array}}
	\newcommand{\ea}{\end{array}}

\renewcommand{\bf}[1]{\mathbf{#1}}

\newcommand{\proj}[1]{\ket{#1}\bra{#1}}

\usepackage[compat=1.1.0]{tikz-feynman}

\newcommand{\eq}[1]{Eq.~\eqref{#1}}
\newcommand{\Eq}[1]{Equation~\eqref{#1}}
\newcommand{\fig}[1]{Fig.~\ref{#1}}

\renewcommand{\neg}{{(-)}}
\newcommand{\sinc}[1]{\text{sinc}\del{#1}}

\renewcommand{\neg}{{(-)}}

\renewcommand{\t}[1]{\text{#1}}

\newcommand{\fs}[1]{\SI{#1}{\femto\second}}
\newcommand{\eV}[1]{\SI{#1}{\electronvolt}}

\makeatletter
\newcommand*{\rom}[1]{\expandafter\@slowromancap\romannumeral #1@}
\makeatother

\begin{document}
	\title{Wavepacket control and simulation protocol for entangled two-photon-absorption of molecules}
	\author{Bing Gu}
	\email{bingg@uci.edu}
	\affiliation{Department of Chemistry \& Department of Physics and Astronomy, University of California, Irvine, California, 92697, USA}
	\author{Daniel Keefer}
		\affiliation{Department of Chemistry \& Department of Physics and Astronomy, University of California, Irvine, California, 92697, USA}
	\author{Shaul Mukamel}
	\email{smukamel@uci.edu}
	\affiliation{Department of Chemistry \& Department of Physics and Astronomy, University of California, Irvine, California, 92697, USA}
	\graphicspath{{pyrazine/figs/}{coop/figs/}}
\begin{abstract}
	Quantum light spectroscopy, providing novel molecular information non-accessible by classical light, necessitates new computational tools when applied for complex molecular systems.
	We introduce two computational protocols for the molecular nuclear wave packet dynamics interacting with an entangled photon pair to produce the entangled two-photon absorption signal. The first  involves summing over transition pathways in a temporal grid defined by two light-matter interaction times accompanied by the field correlation functions of quantum light. 	The signal is obtained by averaging over the two-time distribution characteristic of the entangled photon state.
	 The other protocol involves a Schmidt decomposition of the entangled light and requires summing over the Schmidt modes.
 We demonstrate how photon entanglement can be used to control and manipulate the two-photon excited nuclear wave packets in a displaced harmonic oscillator model.
\end{abstract}
	\maketitle

	\section{Introduction}

 Numerous  novel spectroscopic techniques which exploit  the  variation of photon statistics upon interaction with matter are made possible by quantum light \cite{Mukamel2020, Dorfman2016}.
 Such quantum spectroscopy has been demonstrated, both theoretically and experimentally, to be a powerful technique that can reveal molecular  information not accessible by classical light \cite{Dorfman2016} and can further enhance the signal-to-noise ratio and resolution beyond the classical limit \cite{Li2021b}.   The incoming photon statistics can be employed as a novel control knob for the optical response functions of matter \cite{Roslyak2009, Chen2021b}.   Quantum optical effects such as the Hong-Ou-Mandel two-photon interference \cite{Hong1987} can be used to generate  signals which have no classical analogs \cite{Eshun2021, Kalashnikov2017, Dorfman2021}.

Entangled two-photon absorption (ETPA) has recently attracted considerable attention \cite{Landes2021, Raymer2013b, Oka2020, Terenziani2008, Mollow1968, Schlawin2018a, Li2020c, Landes2021, Eshun2018, Parzuchowski2020, Oka2015, Gu2020, Kang2020a, Gea-Banacloche1989, Guzman2010a, Varnavski2020, Szoke2021, Fei1997, Mollow1968}. {The extent to which two-photon absorption rate can be enhanced by entangled light is under debate. While several ETPA experiments have been reported \cite{Lee2006, Tabakaev2021}, a recent theoretical analysis shows that ETPA events is below the detection level for typical molecular systems in realistic experimental setup \cite{Landes2021}.}  In ETPA, a molecule is promoted from the ground state to an excited state by simultaneously absorbing two,  degenerate or nondegenerate, photons.
This technique drastically differs from classical two-photon absorption.
At low photon fluxes, it
scales linearly, rather than  quadratically with the pump photon intensity \cite{Javanainen1990, Varnavski2020}. This is because the entangled photon pair generated by e.g.  spontaneous parametric downconversion \cite{Couteau2018, Rubin1994, Rubin1996, Keller1997}  is created simultaneously and thus interacts with molecules at the same time. Furthermore, when a narrowband pump is utilized in the twin-photon generation, they exhibit a strong frequency-anticorrelation, i.e., detection of one photon reveals the frequency of its twin within a small uncertainty determined by the pump bandwidth. This energy-time entanglement can be used in spectroscopy to  manipulate the quantum interference among transition pathways, which have been shown to induce classically disallowed collective excitations \cite{Muthukrishnan2004a}, and to probe classically-dark bipolariton states \cite{Gu2020}.
%
Since the ETPA signal depends on the entangled photon pair statistics, the biphoton joint    spectral amplitude can be used to optimize the two-photon excitation process, leading to quantum control by entangled light \cite{Schlawin2017a}.
Early applications have focused on controlling the electronically excited-state populations in model systems with frozen nuclear motion \cite{Schlawin2017a}.
However,  nuclear motions are responsible for reactive dynamics initiated by two-photon absorption, especially for molecules passing through conical intersections with strong vibrational-electronic (vibronic) coupling.

To fully describe the coupled electron-vibration-photonic dynamics,   we  develop a computational framework involving  molecular nuclear  wave packet quantum dynamics interacting with quantum light.
While the response to classical laser pulses can be obtained by solving the \tdse, this is not the case for  quantum light.

 Our simulation protocol, based on  time-dependent perturbation theory for the light-matter interaction, involves summing over all two-photon transition pathways in a two-dimensional temporal grid defined by the two interaction times with the entangled photon pair.
Using a displaced harmonic oscillator model widely used to describe vibronic transitions \cite{Mukamel1995}, we demonstrate how photon entanglement, a novel control knob not available for classical light, can be utilized to control the nuclear wave packet in  electronically excited states.
%
 Three  scenarios are examined whereby the intermediate electronic state is resonant, off-resonant, and, far off-resonant with the incoming photons. We find that the two-photon-excited population is  largest for the resonant case. A linear dependence of the final-state population on the entanglement time is found for short entanglement times. This observation is rationalized by an analytical analysis based on sum-over-states expression.  We further demonstrate that entanglement provides a useful control knob for the two-photon excited wave packets. Modulating the entanglement time has the strongest effect on the shape of created nuclear wave packet in the off-resonant case.






	\section{Theory and Computation}
		\subsection{ The entangled two-photon absorption signal}

We consider a molecule-photon system  described by the Hamiltonian ($\hbar = e = 4\pi \epsilon_0 = 1$) 
	\be
	H = H_\text{M} + H_\text{R} + H_\text{RM}.
	\ee
	 The molecular Hamiltonian 	$ H_\text{M} = T_\text{n} + H_\text{BO}(\bf R)$ represents the  vibronic dynamics
	where $T_\text{n}$ is the nuclear kinetic energy operator,  $H_\text{BO}(\bf R) = \sum_\alpha V_\alpha(\bf R)\proj{\psi_\alpha(\bf R)}$ the adiabatic (Born-Oppenheimer) electronic Hamiltonian with $V_\alpha(\bf R)$ being the $\alpha$th   potential energy surface (PES).
		The radiation Hamiltonian  \be H_\text{R} = \sum_{j = \text{s, i}} \int_0^\infty \dif \omega  \hbar \omega \del{a_j^\dag(\omega_j)a_j(\omega_j) + \half}  \ee
 represents two continua of photon modes, signal and idler, generated by a  SPDC process,
   $H_\text{RM} = \sum_j - \bm \mu \cdot \bf E_j(\bf r)$ is  the light-matter interaction in the electric dipole approximation ,
where $\bf E_j(\bf r) = 	\bf E_j^\pos(\bf r) +	\bf E_j^\neg(\bf r)$ and	\be
	\bf E_j^\pos(\bf r) = i \inth \dif \omega \mc{E}(\omega) \bf e_j a_j(\omega) e^{-i \bf k \cdot \bf r}
	\label{eq:efield}
	\ee
	the electric field operator of the $j$th photon beam with polarization $\bf e_j$ at molecular location $\bf r$, and $\mc{E}(\omega) \equiv \sqrt{\frac{2\pi \omega}{ c n A}}$ with the  beam transversal area $A$, the refractive index $n$, and the speed of light $c$.
	The joint electron-nuclei-photonic space is given by the tensor product of the electron-nuclear Hilbert space  and the photon modes  Fock space. The joint light-matter state is
	\be \ket{\Psi(t)} = \sum_{\alpha, \bf n} \ket{\psi_\alpha(\bf R)} \ket{\chi_{\alpha, \bf n}(t)} \ket{n(\omega) \cdots }
	\ee
	where $\ket{\psi_\alpha(\bf R)}$ is the adiabatic electronic state, $\chi_{\alpha, \bf n}(\bf R, t)$ the nuclear wavefunction at the $\alpha$th electronic state and the photon state is described in the occupation number representation $\ket{\bf n} \equiv \ket{n(\omega) n(\omega') \cdots}$.
	The initial state of the joint light-matter system is
	$ \ket{\psi_0(\bf R)} \ket{\chi_0} \ket{\Phi_0}
	$
	where $\ket{\Phi_0}$ describes the quantum light.
	%
%
%
%
%
	The probability to  arrive at the final electronic state $\ket{\psi_f}$ at time $t$ is given by
	\be
	P(t) = \int \dif \bf R \braket{\Psi(t)| \proj{\psi_f(\bf R), \bf R}\otimes \proj{0}|\Psi(t)} = \int \dif \bf R \abs{\chi_f(\bf R, t)}^2
	\ee
where
\be \chi_f(\bf R, t)  = \bra{ \psi_f(\bf R), \bf R} \otimes \bra{0}\ket{\Psi(t)} \label{eq:112}
\ee  is the nuclear wavefunction  for the final electronic state created by the two-photon excitation process where the photon modes are in the vacuum state $\ket{0}$.   This is obtained by projecting the total state  onto the final electronic state and the photon vacuum after the pump pulse  $t > t_\text{p}$.
 The joint light-matter state at time $t$ is obtained using  time-dependent perturbation theory  in the light-matter interaction 	in the interaction picture of $H_0 = H_\text{M} + H_\text{R}$. To second-order, we have
	\be
	\ket{\tilde{\Psi}(t)} = - \sum_{i, j} \int_{t_0}^t \dif t_2 \int_{t_0}^{t_2} \dif t_1 \del{\bm \mu(t_2) \cdot \bf e_j} \del{\bm \mu(t_1)\cdot \bf e_i} \ket{\psi_0}\ket{\chi_0} { E_j^\pos(t_2) E_i^\pos(t_1)} \ket{\Phi_0}
	\label{eq:111}
	\ee
	where $A(t) = U_0^\dag(t) A U_0(t)$ is the operator $A$ in the interaction picture,  $i,j$ are the beam indices, and $\ket{\tilde{\Psi}(t)} = U^\dag_0(t) \ket{\Psi(t)}$. We consider two incoming beams  and neglect the zero- (no interaction) and first-order (one-photon) processes in \eq{eq:111} as they do not contribute to the two-photon absorption. 	The dipole operator acts in the joint electron-nuclear space
	\be
\bm	\mu = \sum_{\alpha \ne \beta} \int \dif \bf R \braket{\psi_\beta(\bf R)|\bm \mu|\psi_\alpha(\bf R)}_{\bf r} \ket{\psi_\beta(\bf R), \bf R}\bra{\psi_\alpha(\bf R), \bf R}
	\label{eq:dip}
	\ee
	where $\alpha, \beta$ labels the electronic states and $\braket{\cdots }_{\bf r}$ refers to integrating over electronic degrees of freedom. We further  assume that the molecule has no permanent dipole moments so that $\alpha \ne \beta$.
	Inserting \eq{eq:111} into \eq{eq:112} yields
	\be
		\tilde{\chi}_f(\bf R, t) = - \sum_{i, j} \sum_{e} \int_{t_0}^t \dif t_2 \int_{t_0}^{t_2} \dif t_1 V^{(j)}_{f e}(\bf R, t_2)V^{(i)}_{e g}(\bf R, t_1){\chi_0(\bf R)}  \Phi_{ji}(t_2, t_1)
		\label{eq:110}
	\ee
	where $V^{(j)}_{\beta \alpha}(\bf R) = \braket{\psi_\beta(\bf R)| \bm \mu \cdot \bf e_j | \psi_\alpha(\bf R)}_{\bf r}$ is the transition dipole moment  projected on the polarization of $j$th photon beam, $e$ runs over all intermediate electronic surfaces, and
	\be \Phi_{ji}(t_2, t_1) = \braket{0| E_j^\pos(t_2) E_i^\pos(t_1)|\Phi_0}
	\ee
	is the entangled two-photon transition amplitude.
	Transforming back to the Schrodinger picture we obtain
	\begin{equation}
		\begin{split}
	\chi_f(\bf R, t) &= - \sum_{i, j} \sum_{e} \int_{t_0}^t \dif t_2  \int_{t_0}^{t_2} \dif t_1  \xi(t_2, t_1) \\
	 \xi(t_2, t_1) &\equiv U_f(t, t_2) V^{(j)}_{f e}(\bf R)  U_e(t_2, t_1) V^{(i)}_{e g}(\bf R) U_g(t_1, t_0){\chi_0(\bf R)} \Phi_{ji}(t_2, t_1)
	\label{eq:main}
	\end{split}
	\end{equation}
	where $U_\text{M}(t,t') = e^{- i H_\text{M} (t-t')}$ is the molecular free  propagator.
The wave packet $\xi(t_2, t_1)$  represents a single two-photon-absorption event with two light-matter interactions occurring at $t_1$ and $t_2$. This
 reduces to the $\ket{g} \rightarrow \ket{f}$  transition amplitude  if the nuclear motion is frozen.

\begin{figure}[hbt]
	\includegraphics[width=0.2\textwidth]{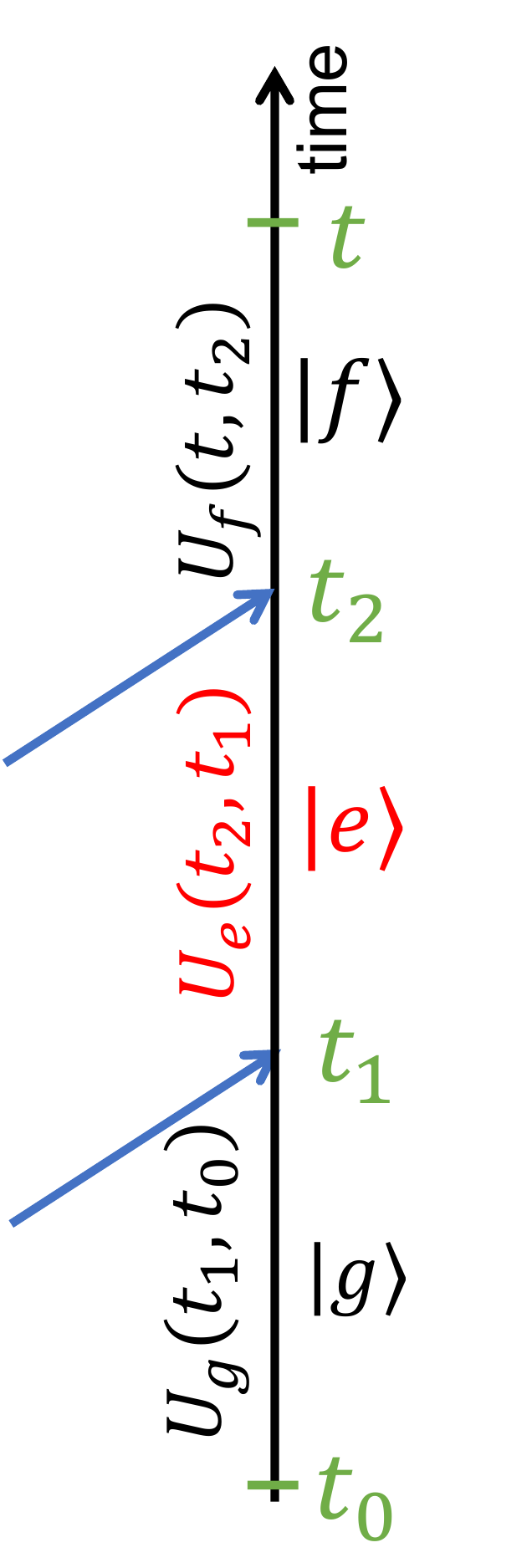}
	\caption{Feynman diagram  representing \eq{eq:main}. $\ket{g}, \ket{e}, \ket{f}$ represent the ground, intermediate, and final electronic states. The $U_\alpha(t, t') = e^{i \del{T_\t{n} + V_\alpha(\bf R)}(t-t')}$ is the free-molecule propagator on the $\alpha$th PES. Between $t_0$ and $t_1$, the molecule remains in the ground state. The first photon, either signal and idler, brings the molecule to the intermediate state, launching nuclear dynamics on the associated PES. }
	\label{fig:loop}
\end{figure}

Using the Feynman diagram  \fig{fig:loop}, \eq{eq:main} can be interpreted as follows: The final nuclear wave packet at time $t$ on the $f$th PES is given by a sum over of all possible transition pathways. Each  pathway contains two dipole interaction times $t_1$ and $t_2$, where the molecule undergoes a transition between  electronic states. Between $t_0$ and $t_1$, the molecule remains at the ground state. At  $t_1$, the molecule makes a transition to the $e$th electronic state, launching a nuclear wave packet dynamics until it interacts with the second photon  at $t_2$, which brings it to the final $f$-PES.  Nuclear dynamics then takes place between $t_2$ and the final time $t$ on this PES.
 Each two-photon transition pathway depends on the two-photon transition amplitude $\Phi_{ji}(t_2, t_1)$, whose modulus squared gives the probability of detecting the $i$-photon at time $t_1$ and $j$-photon at $t_2$.  The molecule thus serves as a photodetector \cite{Glauber1963}.
	The dependence of the ETPA signal on the photon statistics allows to control the two-photon excited nuclear wave packet on the excited-state PES by shaping the two-photon wave function.

%
%


	\subsection{Energy-Time entangled light} \label{sec:light}
\begin{figure}[hbt]
		\includegraphics[width=0.8\textwidth]{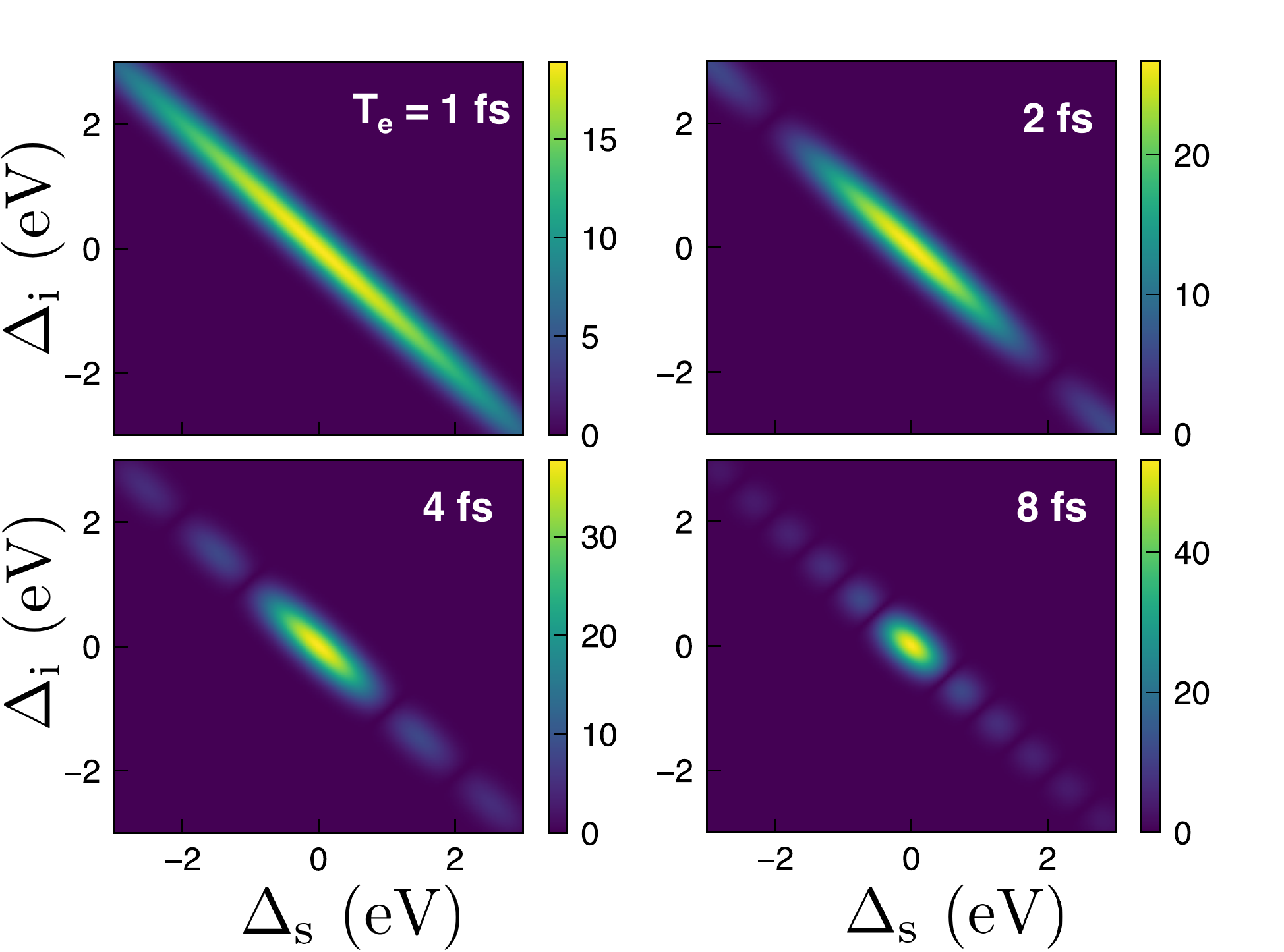}
		\caption{The joint spectral amplitude $\abs{J(\Delta_\text{s}, \Delta_\text{i})}$ of the quantum light [\eq{eq:116}] for different entanglement times $T_\t{e}$. Here $\sigma_\text{p} = \eV{0.2}, \bar{T} = 0$. }
		\label{fig:light}
	\end{figure}

	 The two-photon wavefunction is described by the joint spectral amplitude (JSA) of the entangled photon pair.
	Here we employ the twin-photon, signal and idler, state, generated by a type-II SPDC process \cite{Rubin1994, Rubin1996, Fei1997},
	\be
	\ket{\Phi} = \iint_0^\infty \dif \omega_\text{s} \dif \omega_\text{i} J(\omega_\text{s}, \omega_\text{i})a^\dag_\text{s}(\omega_\text{s}) a^\dag_\text{i}(\omega_\text{i}) \ket{0}
	\ee
	where the JSA $J(\omega_\text{s}, \omega_\text{i})$ is the  amplitude of detecting the signal photon with frequency $\omega_\text{s}$ and idler photon with frequency $\omega_\text{i}$, and $a_j(\omega)$ ($a^\dag_j(\omega)$) annihilates (creates) a $j$-photon with frequency $\omega$ satisfying the boson commutation relation $\sbr{a_j(\omega), a^\dag_{j'}(\omega')} = \delta_{jj'} \delta(\omega - \omega')$.
	We focus on the frequency anti-correlation of the entangled photons, and suppress the spatial degrees of freedom \cite{Walborn2010, Rubin1996}.
	For a  Gaussian pump with central frequency $\bar{\omega}_\t{p}$ and bandwidth $\sigma_\text{p}$, the JSA reads
	\be
	J(\omega_1, \omega_2) = \mc{N}  \exp\del{-\frac{\del{\omega_1 + \omega_2 - \bar{\omega}_\text{p}}^2}{4 \sigma^2_\text{p}}} \sinc{\frac{\Delta k L}{2} }
	\label{eq:jsa}
	\ee
	where $L$ is the crystal length, $\Delta k = k_\text{s}(\omega_\text{s}) + k_\text{i}(\omega_\text{i}) - k_\text{p}(\omega_\text{s} + \omega_\t{i})$ is the wavenumber mismatch and $k_j = n_j(\omega_j)\omega_j/c$, $\mc{N}$ is the normalization factor ensuring $\iint \dif\omega_\t{s} \dif \omega_\t{i} \abs{J(\omega_\t{s}, \omega_\t{i})}^2 = 1$, and $\sinc{x} = \frac{\sin(x)}{x}$.
	We focus on the degenerate state where the central frequencies for the signal and idler photons are identical $ \bar{\omega}_\text{s} = \bar{\omega}_\text{i} = \half \bar{\omega}_\text{p}$.  Taylor expansion of the wave number gives
	$k_j(\omega_j) \approx k_j(\bar{\omega}_j) + \Delta_j/v_{j}$ where $\Delta_j = \omega_j - \bar{\omega}_j$ and $v_{j} \equiv \od{\omega_j}{k_j}$ is the group velocity of $j$th beam. Under the phase matching condition
	$
	k_\text{s}(\bar{\omega}_\t{s}) +  k_\t{i}(\bar{\omega}_\t{i}) -  k_\text{p}(\bar{\omega}_\text{p}) = 0
	$, the JSA becomes
	\be
	J(\Delta_\t{s}, \Delta_\t{i})
	= \mc{N} \exp\del{-\frac{\del{\Delta_\text{s} + \Delta_\t{i}}^2}{ 4 \sigma^2_\text{p}} } \sinc{ \half \bar{T} \del{\Delta_s + \Delta_i} +  \frac{1}{2} \del{\Delta_\text{s} - \Delta_\text{i}} T_\text{e} }
	\label{eq:116}
	\ee
	where $T_\text{e} = \half \del{\frac{L}{v_{\text{s}}} - \frac{L}{v_\text{i}}}$ is the entanglement time characterizing the difference between the arrival times of the photon pair and
	$
	\bar{T} = \half \del{\frac{L}{v_{\text{s}}} + \frac{L}{v_\text{i}}} - \frac{L}{v_\text{p}}
	$ is the travel time difference between the biphoton and the pump inside the nonlinear crystal.
 The twin-photon JSA  for different entanglement times are shown in \fig{fig:light}. Each photon bandwidth is controlled by the entanglement time with shorter $T_\t{e}$ leading to broader bandwidth, whereas the sum frequency of the signal and idler beams are narrowly distributed independent of the bandwidth of individual photon.

	The two-photon detection amplitude is given by
	\be
	\Phi_{\t{is}}(t_2,t_1) = \iint_0^\infty \dif \omega_\t{s} \dif \omega_\t{i}  \mc{E}\del{\omega_\t{i}}\mc{E}\del{\omega_\t{s}} e^{-i\omega_\t{i} t_2 -i \omega_\t{s} t_1} f(\omega_\text{s}, \omega_\text{i})
	\label{eq:amp}
	\ee
	Invoking the narrowband limit $\mc{E}(\omega_j) \approx  \mc{E}(\bar{\omega}_j)$, changing the variables to $\Delta_j$ and extending the integration range to $(-\infty, \infty)$, yields
	\be
	\Phi_{\t{is}}(t_2,t_1) =  \del{2\pi}^2 \mc{E}\del{\bar{\omega}_\t{i}}\mc{E}\del{\bar{\omega}_\t{s}} \exp\del{-i\bar{\omega}_\t{i} t_2 -i \bar{\omega}_\t{s} t_1} J(t_1, t_2)
	\ee
	where $J(t_1, t_2) = \iint_{-\infty}^{\infty} \frac{\dif \Delta_\text{s}  \dif \Delta_\text{i}}{\del{2\pi}^2} J(\Delta_\t{s}, \Delta_\t{i}) e^{-i\Delta_\t{i} t_2 -i \Delta_\t{s} t_1} $ is the joint temporal amplitude of the twin-photons.
	For $\bar{T} = 0$,
	\be
	J(t_1, t_2)  = \sqrt{{\sigma_\text{p}}/{T_e}} \del{2\pi}^{-5/4} e^{-\sigma_\text{p}^2 \del{t_1 + t_2}^2/4} \Pi\del{\frac{{t_1-t_2}}{2T_e}}
	\label{eq:jta}
	\ee
	where $\Pi(x) = 1 $ for $\abs{x} < \half$ and $0$ otherwise.

\subsection{ Simulation Protocol Based on a Time Grid}

\Eq{eq:main} suggests the following simulation protocol: we first sample $\del{t_2, t_1}$ on a two-dimensional triangular grid with $t_2$ ranging from $t_0$ to the final time $t$, and $t_1$  samples $t_0$ to $t_2$; for each $\del{t_1, t_2}$, we compute the nuclear wave packet $\xi(t_2, t_1)$ by a wave packet dynamics solver. The final wave packet can then be obtained by a sum over all  $\xi\del{t_2, t_1}$.

The final nuclear wave packet \eq{eq:main} is simulated as follows (for brevity, we assume a single intermediate electronic state $e$):
\begin{enumerate}
	\item  Set the initial wave packet to $ \chi_e(\bf R) = \mu_{eg}(\bf R) \chi_0(\bf R)$
	\item Propagate the wave packet on the $e$th PES for time interval $\tau$  leading to $\chi_e(\bf R, \tau) = U_\text{M}(\tau) \mu_{eg}(\bf R) \chi_0(\bf R) $.
\item Using \eq{eq:amp}, perform the following integration to obtain an auxiliary wavepacket  $\zeta(\bf R, t)$,
		\be
	\zeta(\bf R, \tau_2) =  \int_0^{\tau_2} \dif \tau \Phi(\tau_2 + t_0,\tau_2-\tau+t_0)  \chi_e(\bf R, \tau)
	\ee
	\item  For each $\tau_2 = t - t_0 - t_2$, apply the dipole operator to $\zeta(\bf R, \tau_2)$ that brings the molecule to the final electronic state,
		$ \mu_{fe}(\bf R) \zeta(\bf R, \tau_2)$, and propagate the  wave packet for $t - t_0 - \tau_2$ on the final PES.
	Summing up all possible $\tau_2$ leads to
	\be \chi_f(\bf R, T=t-t_0) = \int_{0}^{T}\dif \tau_2 U_\text{M}(t - t_0 - \tau_2)  \mu_{fe}(\bf R)  \zeta(\bf R, \tau_2)
	\ee

\end{enumerate}
%
%
The molecular propagator is computed using a wave packet dynamics  on a single PES \cite{Kosloff1988}.
The  second-order split-operator method is employed for adiabatic wave packet dynamics on a single PES $V_\alpha(\bf R)$.  A Trotter decomposition of the propagator is employed
\be U_\alpha(\delta t) = e^{-i V_\alpha \delta t/2} e^{-i T_\text{n} \delta t} e^{-i V_\alpha \delta t/2} + \mc{O}((\delta t)^3)
\ee
for a short time interval $\delta t$ and  fast Fourier transform switching the wavefunction between the coordinate and momentum space.


	\section{Results and Discussion}

\begin{figure}[hbt]
	\includegraphics[width=0.8\textwidth]{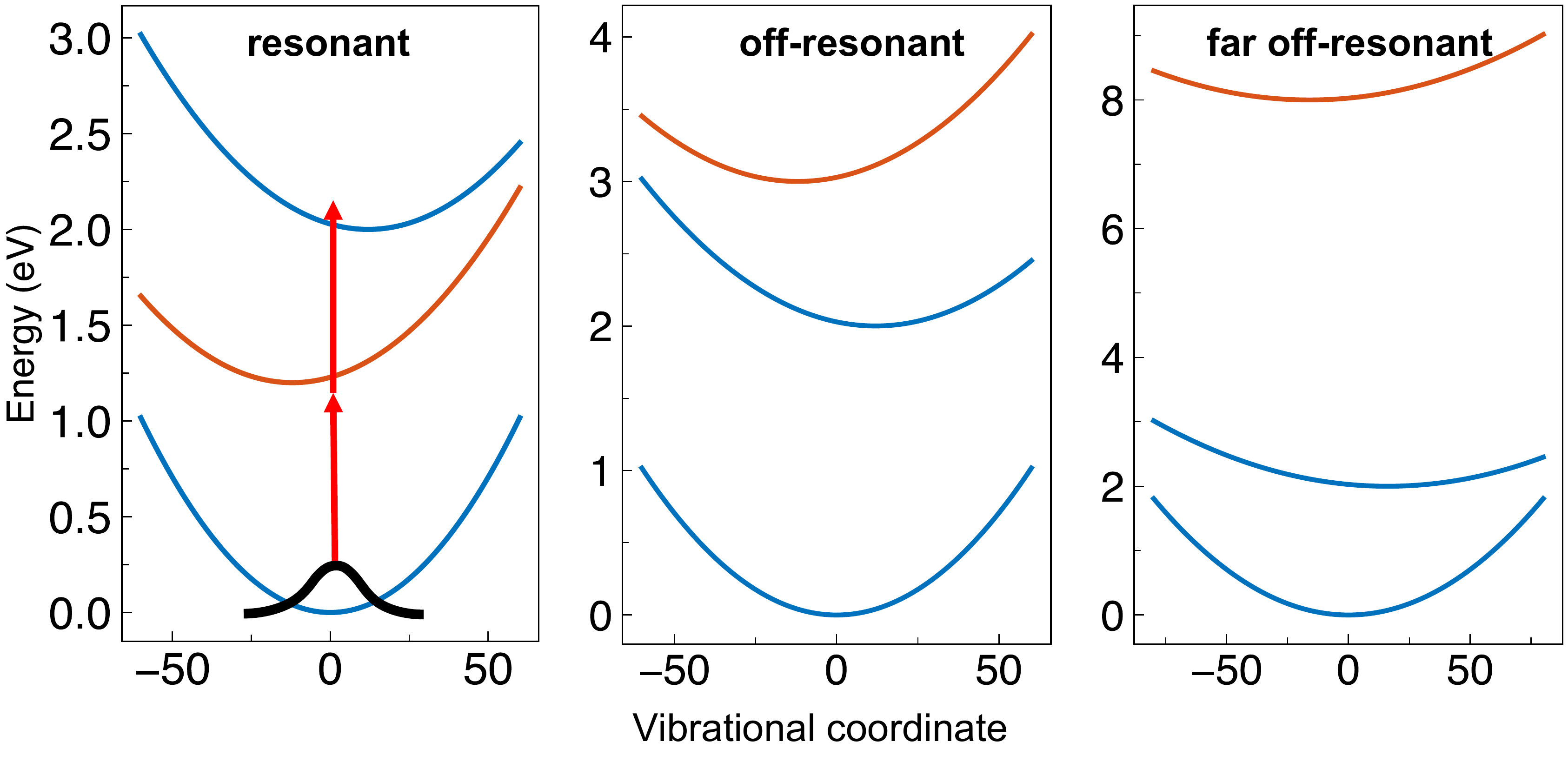}
	\caption{Potential energy surfaces of the displaced harmonic oscillator model corresponding to three different cases where the intermediate electronic state is resonant, off-resonant, and far off-resonant with the incoming photons. Correspondingly, $E_e = 1.2, 3, \eV{8}$. }
	\label{fig:model}
\end{figure}

Simulations were carried out for a three-state displaced harmonic oscillator model with a single nuclear coordinate $x$ and corresponding momentum $p$. The PESs depicted in \fig{fig:model} are given by
	\be
	V_\alpha(x) = \frac{p^2}{2} + \frac{\omega_\alpha^2}{2} \del{x - d_\alpha}^2 + E_\alpha
	\ee
where $\alpha = \set{g, e, f}$ referring to the ground, intermediate, and final electronic states, respectively, and $d_\alpha$ is the displacement, and $E_\alpha$ is the zero-phonon line. By tuning the energy $E_e$, we cover  three  scenarios whereby the intermediate PES is resonant, off-resonant and far off-resonant with respect to the incoming photons. Other parameters are  $E_g=0, E_f = \eV{2},  d_g = 0, d_e = -d_f = -10 \si{\bohr}$.


\begin{figure}[hbt]
	\includegraphics[width=0.4\textwidth]{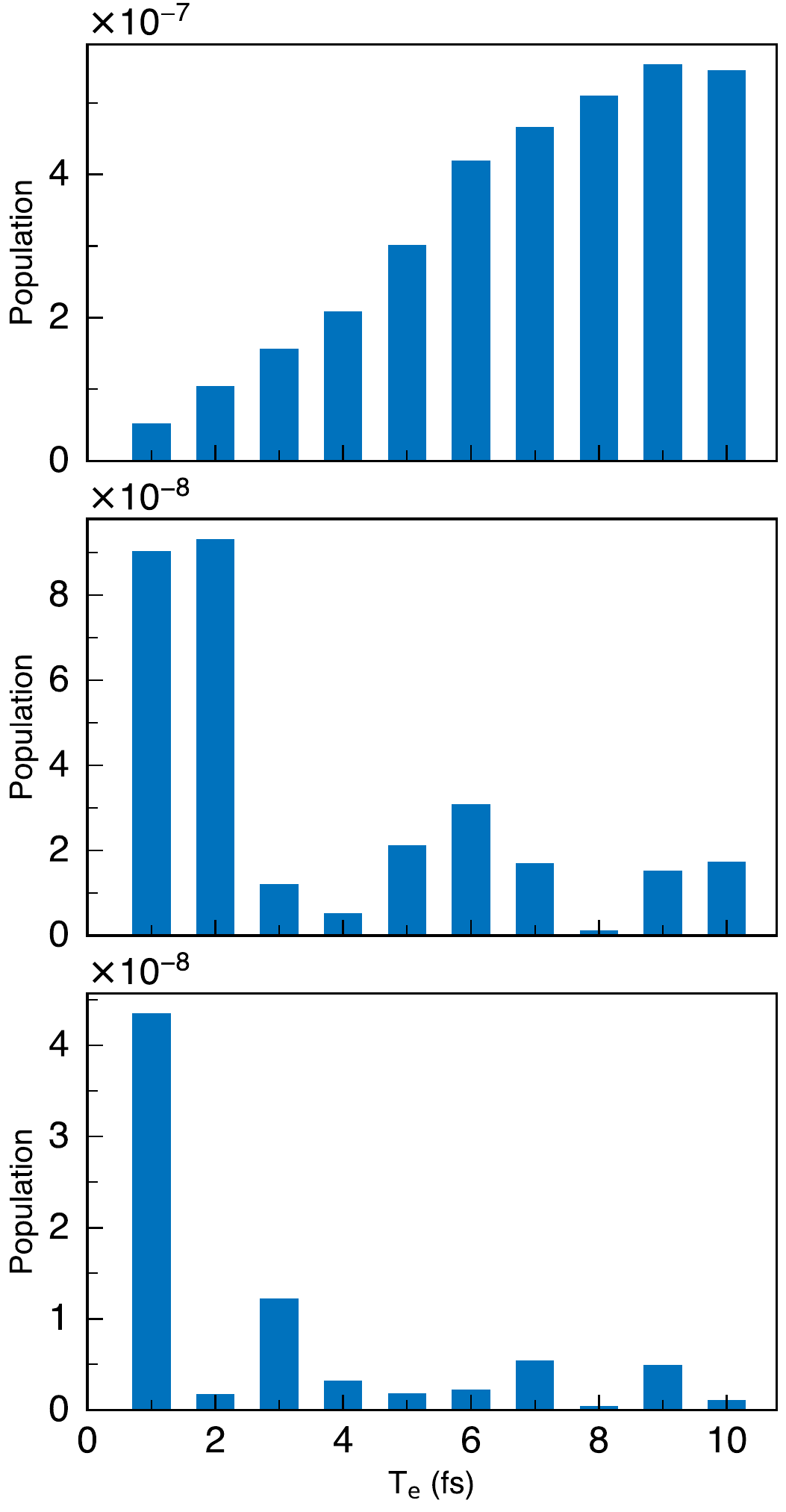}
	\caption{Dependence of the entangled two-photon-excited population on the entanglement time. The parameters read $A = 1 \si{\micro\meter \squared}, n = 1, \bar{\omega}_\t{p} = \eV{2.4}, \sigma_\text{p} = \eV{0.2}$. }
	\label{fig:population}
\end{figure}

\begin{figure}[hbt]
	\includegraphics[width=0.8\textwidth]{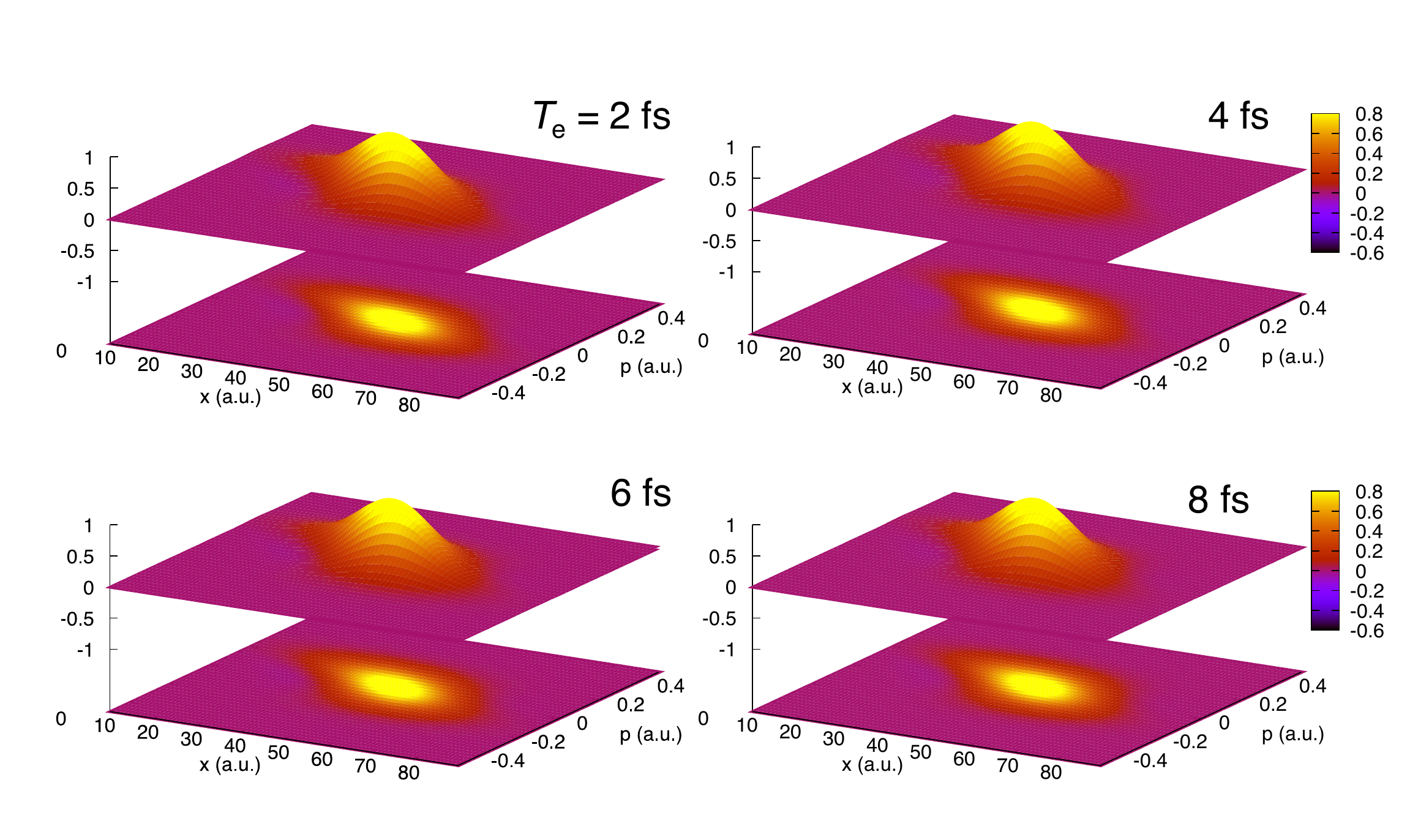}
	\caption{  Phase space representation  of entangled two-photon-excited wavepackets for the displaced harmonic oscillator model with a resonant intermediate electronic state for various entanglement times as indicated. }
	\label{fig:resonant}
\end{figure}

\begin{figure}[hbt]
	\includegraphics[width=0.8\textwidth]{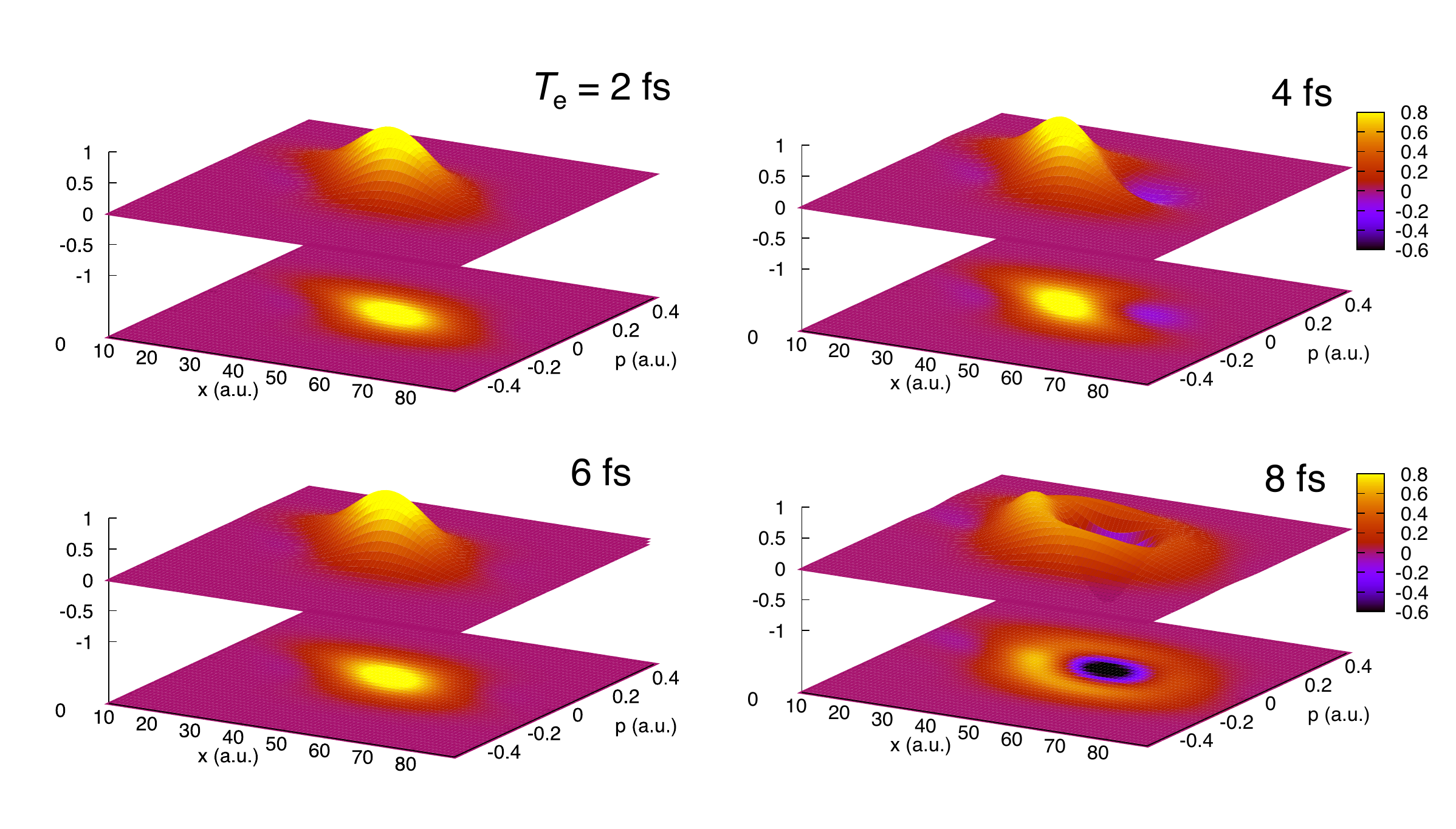}
	\caption{ Same as \fig{fig:resonant} but for off-resonant intermediate electronic state.
	 }
 	\label{fig:wp2}
\end{figure}

\begin{figure}[hbt]
	\includegraphics[width=0.8\textwidth]{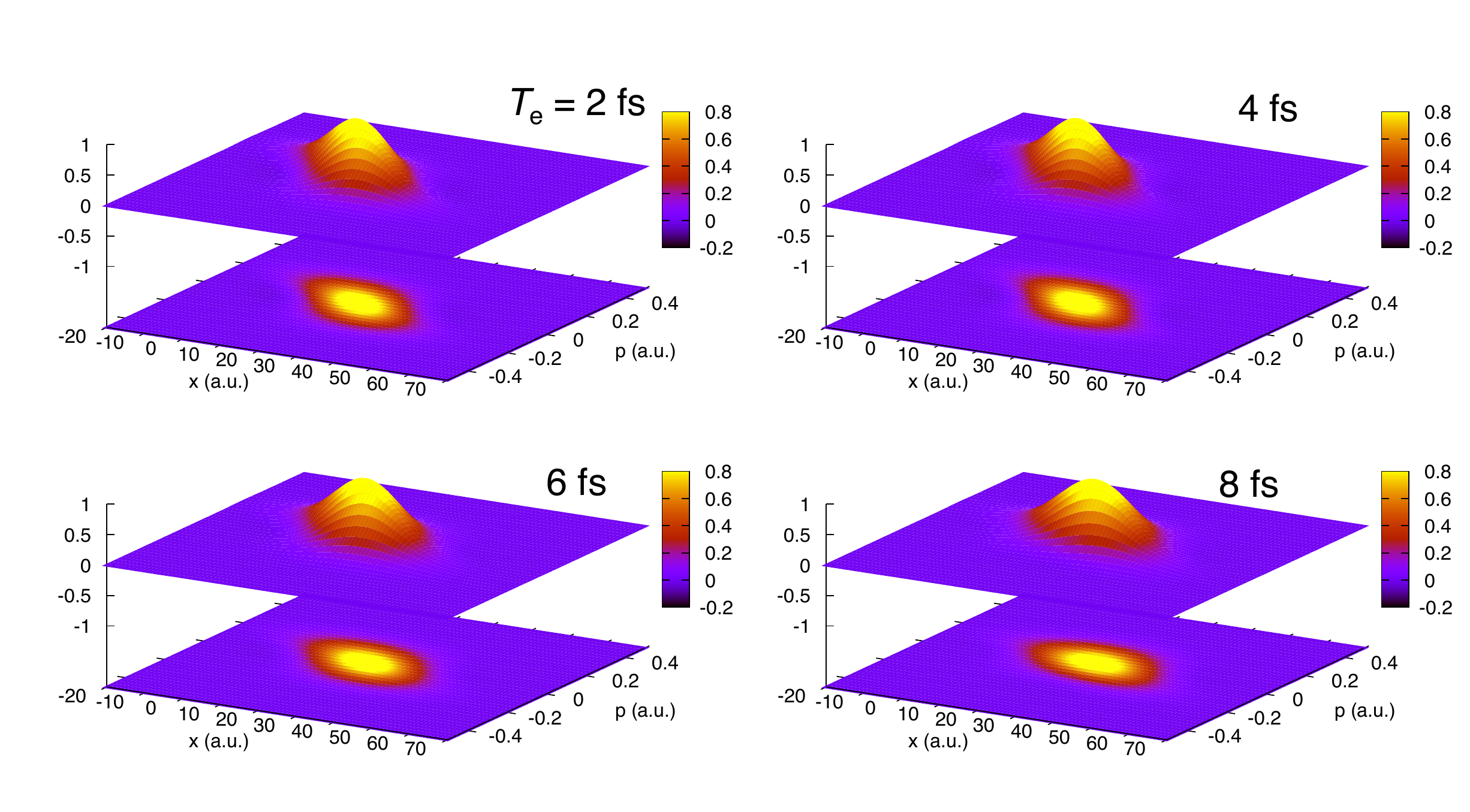}
	\caption{ Same as \fig{fig:resonant} but for far off-resonant intermediate electronic state.
	}\label{fig:wp3}
\end{figure}

	\subsection{  Sum-over-states expansion of the molecular response}
	We explore the variation of the electronic populations and wave packets with the entanglement time by employing the sum-over-states expression for the molecular response in the vibronic eigenstates of the molecular Hamiltonian $H_\text{M}$. Let $\ket{\alpha \nu}$ denote the vibronic states associated with  $\alpha$th electronic state with eigenenergies $\omega_{\alpha \nu}$, the molecular propagator and the interaction picture dipole operator is
	$
	U_\text{M}(t) =\sum_{\alpha} \sum_\nu e^{- i \omega_{\alpha \nu} t} \proj{\alpha \nu}
	$ and $V^{(j)}(t) = \sum_{\beta \nu', \alpha \nu} V^{(j)}_{\beta \nu', \alpha \nu} e^{i \omega_{\beta \nu', \alpha \nu} t} \ket{\beta \nu'}\bra{\alpha \nu}   $ with   $\omega_{\beta \nu', \alpha \nu} =  \omega_{\beta \nu'} -  \omega_{ \alpha \nu}$.
Inserting these into \eq{eq:main} yields
$ \ket{\chi_f( t)} = \sum_\nu A_{f\nu, g0}(t) \ket{f\nu} $
where \be
T_{f\nu, g0}(t) = \del{2\pi}^2 \mc{E}\del{\bar{\omega}_\t{i}}\mc{E}\del{\bar{\omega}_\t{s}}
\sum_{e, \nu'} \mu_{f\nu, e\nu'}^{(\t{i})} \mu_{e\nu', g0}^{(\t{s})}
\int_{t_0}^t \dif t_2 e^{i\del{\omega_{f\nu, e\nu'} - \bar{\omega}_\t{i}} t_2}\int_{t_0}^{t_2} \dif t_1  e^{i\del{\omega_{e\nu', g0} - \bar{\omega}_\text{s}} t_1} J(t_1, t_2)
\ee
is  the transition amplitude from the ground vibrational state in the ground electronic state PES $\ket{g0}$ to the vibronic state $\ket{f\nu}$.
Using  \eq{eq:jta} and taking $t \rightarrow \infty, t_0 \rightarrow -\infty$ yields
\be
A_{f\nu, g0} = \del{2\pi}^{3/4} \sqrt{\frac{\pi}{\sigma_\text{p} T_\t{e}}}\mc{E}\del{\bar{\omega}_\t{i}}\mc{E}\del{\bar{\omega}_\t{s}}
\exp\del{- \frac{ \del{\omega_{f\nu, g0} - \bar{\omega}_\t{p}}^2}{4\sigma_\text{p}^2}}
\sum_{\nu'} \mu_{f\nu, e\nu'}^{(\t{i})} \mu_{e\nu', g0}^{(\t{s})}
 \frac{e^{i \Delta_{\nu'} T_e} - 1}{ i {\Delta_{\nu'}}} + \del{\text{s} \leftrightarrow \t{i}}
 \label{eq:sos}
\ee

where $\nu'$ runs over the vibrational eigenstates in $e$th PES, $\Delta_{\nu'} = \half \omega_{f\nu, g0} - \omega_{e\nu', g0}$ and $\mu_{\beta \nu', \alpha \nu} =  \braket{\beta \nu' | \mu | \alpha \nu}$ is the transition dipole moment between vibronic states.




\fig{fig:population} depicts the two-photon-excited population as a function of the entanglement time.  The largest excited population occurs for the resonant case (upper panel), as reflected in the detuning factor $1/\Delta_{\nu'}$ in \eq{eq:sos}. Interestingly, the population grows, roughly linearly, with $T_\t{e}$ at short entanglement times. To rationalize this observation, we isolate the $T_\t{e}$-dependent factor in \eq{eq:sos} $g(T_\t{e}) = \frac{1}{\sqrt{T_\t{e}}}   \frac{e^{i \Delta_{\nu'} T_e} - 1}{ i {\Delta_{\nu'}}}$ and we use $e^{x} \approx 1 + x$
\be
P \propto \abs{g(T_\text{e})}^2  \approx T_\t{e}.
\label{eq:114}
\ee
Thus, for resonant intermediate states and short entanglement times, the two-photon-excited population grows linearly with $T_\t{e}$. After the first photon interacts with the molecules, transient population is built in the intermediate state. This  population grows for a short period of time until the second photon arrives. The time window is bounded by the entanglement time of the quantum light, whereas for classical light there is so such restriction.  This linear increase only exists at very short entanglement times below $T_\t{e} = \fs{10}$.

For off-resonant and far off-resonant intermediate states (middle and lower panels of \fig{fig:population}), the  two-photon-excited population shows a nonlinear dependence on the entanglement time. In both cases the largest population occurs at short entanglement times with the population for the off-resonant case larger than the far off-resonant case, as expected.

Apart from controlling the electronic populations, the JSA may also be used to manipulate the nuclear wave packet. Figs. \ref{fig:resonant}, \ref{fig:wp2}, and \ref{fig:wp3} show the phase-space Wigner representation of the nuclear wave packets prepared by entangled light with various entanglement times at $t = \fs{20}$ for the resonant, off-resonant, and far off-resonant cases, respectively. The Wigner spectrogram transforms the wavepacket in coordinate space as
$
\chi_\t{W}(x, p, t) = \intf \dif y \chi_f\del{x + \frac{y}{2} } \chi_f^*\del{x - \frac{y}{2}} e^{i p y}
$.
We see that the  nuclear wavepacket is most sensitive to the entanglement time in the off-resonant case  with minor variations otherwise.  The nuclear wave packet depends on both the amplitude and phase of the transition amplitude to a vibronic state in the $f$th electronic state $A_{f\nu}$. For the resonant case, \eq{eq:114} implies that for short entanglement times, the relative phase between vibrational states in the $f$-PES does not depend on $T_\text{e}$. For the off-resonant case, the $g(T_\t{e})$ will show an oscillatory behavior with $T_\text{e}$ and  the relation between $A_{f\nu}$ and the vibrational state $\nu$ will  strongly depend on $T_\t{e}$ thus leading to a considerable change in the wavepacket.

\subsection{Simulation Protocol Based on the Schmidt decomposition}
Sampling the wavepackets on the two-dimensional time grid is numerically expensive. We now present an alternative  simulation protocol for the ETPA, which employs the Schmidt decomposition of the entangled light \cite{Law2004} and replaces the time grid sampling by a summation over Schmidt modes. The entangled light can be expanded in Schmidt modes.  Each pair of modes leads to a transition amplitude, and summing over all contributing Schmidt modes leads to the final signal. The photon-pair entanglement is then reflected in the quantum interference among Schmidt modes.

\subsubsection{ Schmidt decomposition of quantum light} \label{app:decompose}
With the Schmidt decomposition for the JSA  \cite{Law2004, Raymer2019},
$
J(\omega_\t{s}, \omega_\t{i}) = \sum_{n}  \sqrt{\lambda_n} \phi_n\del{\omega_\t{s}} \varphi_n \del{\omega_\text{i}},
$
the two-photon wavefunction can then be expressed by
\be
\Phi_{\t{is}}(t_2, t_1) =  \del{2\pi}^2 \mc{E}\del{\bar{\omega}_\t{i}}\mc{E}\del{\bar{\omega}_\t{s}} \sum_n \sqrt{\lambda_n} \phi_n(t_1) \varphi_n(t_2)
\label{eq:113}
\ee
where $ \phi_n(t) = \intf \frac{\dif \omega}{2\pi} \phi_n(\omega) e^{i \omega t} $
and $\varphi_n(t)$ are the temporal modes, i.e., Schmidt modes Fourier transformed to the time domain, $\phi_n(\omega_\t{s})$ and $\varphi_n(\omega_\t{i})$ are Schmidt modes, that are, respectively, the eigenstates of the reduced density matrices of the signal and idler photons with $\lambda_n$ the corresponding eigenvalues.

Inserting \eq{eq:113} into \eq{eq:main} leads to
\be
\begin{split}
	\chi_f(\bf R, t) &=   \del{2\pi}^2 \mc{E}\del{\bar{\omega}_\t{i}}\mc{E}\del{\bar{\omega}_\t{s}}  \sum_n \sqrt{\lambda_n} \kappa_n(\bf R, t) \\
	\kappa_n(\bf R, t) &\equiv  \sum_e \int_{t_0}^t \dif t_2 \int_{t_0}^{t_2}\dif t_1 U_{f}(t,t_2) V^\dag_{fe}(\bf R) U_{e}(t_2, t_1) V^\dag_{eg}(\bf R)  {\chi_0(\bf R)} \phi_n(t_1) \varphi_n(t_2)
\end{split}
\ee
$\kappa_n(\bf R, t)$ is the two-photon-excited nuclear wavepacket with the $n$th pair of Schmidt modes.
%
An important observation is that the $\kappa_n$ can be simply simulated by solving \tdse\ in the presence of two classical pulses with electric field $\mc{E}_\t{s}(t) = \phi_n(t), \mc{E}_\t{i}(t) = \varphi_n(t)$.
Therefore, instead of using a temporal grid
one can simply solve  the \tdse\  to compute the ETPA signal.

However, in classical two-photon absorption, there are additional second-order transition pathways corresponding to absorbing two photons from a single beam, which does not exist in quantum light.
This becomes clear in the classical TPA expression
\be
\chi_f(\bf R, t) \propto \int_{t_0}^t \dif t_2 \int_{t_0}^{t_2}\dif t_1 V_{fe}(t_2) V_{eg}(t_1) \chi_0(\bf R) \del{\mc{E}_\t{s}(t_2)\mc{E}_\t{s}(t_1) + \mc{E}_\t{i}(t_2) \mc{E}_\t{s}(t_1) + \mc{E}_\t{s}(t_2)\mc{E}_\t{i}(t_1) + \mc{E}_\t{i}(t_2)\mc{E}_\t{i}(t_1) }
\ee
where the additional terms are associated with $\mc{E}_\t{i}(t_2)\mc{E}_\t{i}(t_1)$ and  $\mc{E}_\t{s}(t_2)\mc{E}_\t{s}(t_1)$.
These have the same order as the desired ones absorbing signal and idler photons together, and thus cannot be eliminated by  weakening the field.

\subsubsection{Selecting pathways by phase cycling}

To remove the undesired transition pathways, we can employ a phase cycling protocol \cite{Tan2008, Cho2018}. Phase cycling selectively extracts pathways by applying phases to the pulses,
\be
\mc{E}_j(t) \rightarrow e^{i\theta_j} \mc{E}_j(t)
\ee
for $j = \text{s, i}$. 
It exploits the fact that different pathways respond differently to the phase change.

\begin{table}[ht]
	\begin{tabular}{c|cc|cccc}
		\hline
		& $\theta_\t{s}$ & $\theta_\t{i}$ & ss & ii & si & is \\
		\rom{1} & 0 & $\frac{\pi}{2}$ & 1 & -1 & i & i \\
		\rom{2} &  $\frac{\pi}{2}$ & 0 & -1 & 1 & i & i \\
		\hline
	\end{tabular}
	\caption{Phase cycling protocol to remove additional pathways. The final signal $S = \frac{1}{2i} \del{S_\rom{1} + S_{\rom{2}}}$. The four pathways are labeled by the signal (s) and idler (i) photons interacting with the molecule in the given order.}
	\label{tab:pc}
\end{table}

A phase cycling protocol that eliminates the two-photon transition pathways ss and ii  is shown in Table \ref{tab:pc}.

This protocol with Schmidt modes can be very efficient if the entangled light can be described by a limited number of Schmidt modes \cite{Eberly2006}.
	\section{Conclusions}

 We have presented a computational protocol for the entangled two-photon absorption signal in molecules which takes the nuclear quantum dynamics into account. It involves summing over all transition pathways determined by two light-matter interaction times.  Using a displaced harmonic oscillator model, we have demonstrated how entangled light can be used to manipulate the two-photon-excitation process. Both electronic populations and nuclear wave packets strongly depend on the entanglement time. This protocol applies to any joint spectral amplitude of the entangled light, and thus can be applied for various sources of quantum light from e.g. cascaded emission as well.   We have also outlined an alternative protocol based on the Schmidt decomposition of the entangled light, which can be very efficient if the entangled light can be described by a limited number of Schmidt modes (weak entanglement).
%
Our protocols allows the simulation of quantum light spectroscopy of complex molecular systems fully accounting for the coupled electronic-nuclear-photonic motion. 	 Advances in pulse shaping techniques may allow a complete control of the JSA by varying the parameters other than entanglement time \cite{Schlawin2017a}.

\begin{acknowledgments}
	We thank Dr. Feng Chen for inspiring discussions.
	B.G. and S.M. are supported by the National Science Foundation Grant CHE-1953045 and by the U.S. Department of Energy, Office of Science, Office of Basic Energy Sciences under Award DE-SC0020168. D.K.  gratefully  acknowledges  support  from  the  Alexander  von  Humboldt  foundation  through  the  Feodor Lynen  program.
\end{acknowledgments}

	\appendix

\bibliography{../../qchem,../../optics,../../cavity}


\end{document}